\documentclass[aps,twocolumn,amsmath,showkeys,showpacs,prl]{revtex4-1}
\usepackage{dcolumn}
\usepackage{bm}
\usepackage[T1]{fontenc}
\usepackage{amsmath}
\usepackage{graphicx}
\usepackage{tabularx}

\usepackage{hyperref}
\hypersetup{colorlinks=true, linkcolor=blue, citecolor=blue, urlcolor=blue, pdftitle={Direct magnetization-polarization coupling in BaCuF$_4$ fluoride}, pdfauthor={A. C. Garcia-Castro}}

\begin{document}
\title{Direct magnetization-polarization coupling in BaCuF$_4$ fluoride}
\author{A. C. Garcia-Castro$^{1,2}$}
\email{a.c.garcia.castro@gmail.com}
\author{W. Ibarra-Hernandez$^{3,4}$}
\author{Eric Bousquet$^{1}$}
\author{Aldo H. Romero$^{3,4}$}
\email{alromero@mail.wvu.edu}
\affiliation{$^1$Physique Th\'eorique des Mat\'eriaux, CESAM, Universit\'e de Li\`ege, B-4000 Sart-Tilman, Belgium}
\affiliation{$^2$Department of Physics, Universidad Industrial de Santander, Cra. 27 Cll. 9, Bucaramanga, Colombia}
\affiliation{$^3$Department of Physics and Astronomy, West Virginia University, WV-26506-6315, Morgantown, USA}
\affiliation{$^4$Facultad de Ingenier\'ia-BUAP, Apartado Postal J-39, Puebla, Pue. 72570, M\'exico}

\begin{abstract} 
Despite oxides and some fluorides perovskites have emerged as prototypes of multiferroic and magnetoelectric materials, they have not impacted real devices. Unfortunately, their working temperatures are very low and the magnetoelectric coupling has been reported to be rather small. 
Herewith, we report from first-principles calculations an ideal magnetization reversal through polarization switching in BaCuF$_4$ which, according to our results, could be achieved close to room temperature.
We also show that this ideal coupling is driven by a soft mode that combines both, polarization and octahedral rotation. The later being directly coupled to the weak ferromagnetism of BaCuF$_4$. 
This, added to its strong Jahn-Teller distortion and its orbital ordering, makes this material as a very appealing prototype for crystals in the $ABX_4$ family in multifunctional applications.
\end{abstract}

\maketitle

The search for materials that own ferroelectricity, magnetization, and orbital ordering with a large coupling between those properties, has become one of the most active research fields in condensed matter physics. Since last decade, it has received a vigorous interest by many different research groups due to their high potential in new technologies where multifunctionalities are required.
Among these couplings, magnetoelectricity promises to reduce the computer memory energy consumption, to improve magnetic field sensors or to be used for spintronic applications \cite{Fiebig2016}.
The magnetoelectric crystals, however, suffer from their scarcity, their small response and their low functioning temperature. 
In spite of the great improvements in identifying and understanding the underneath mechanism of magnetoelectricity, 
finding new room-temperature candidates has been difficult and their number is quite scarce.
One of possible solution is to create those materials as composites but magnetoelectric single crystals are still very rare at the present stage of knowledge in this research field \cite{Fiebig2016}.
Another solution to find good magnetoelectric single crystals has been to identify new ferroelectric materials, where the ferroelectric (FE) ordering can be coupled to the magnetization \cite{PhysRevLett.106.107204, bousquet2008}.  
To that end, novel stoichiometries and compositions have been investigated and a promising approach is to look for layered perovskite materials where the octahedra rotations are linked to the polarization through improper-like couplings~\cite{benedek2015, young2015, bousquet2008, PhysRevLett.106.107204}. 
Indeed, Ruddlesden-Popper, Aurivillius, and Dion-Jacobson \cite{benedek2015} phases have been shown to be good candidates to exhibit the coupling between polarization and magnetism. 
The proof of concept has been shown theoretically by Benedek and Fennie in Ca$_3$Mn$_2$O$_7$ \cite{PhysRevLett.106.107204} where the coupling mechanism with magnetism relies on the improper origin of the polarization that indirectly couples with the magnetization.
Unfortunately, the experimental efforts to verify this prediction have, so far, shown that the electric polarization in Ca$_3$Mn$_2$O$_7$ cannot be switched which seems to indicate that this crystal might not be the best candidate for magnetization reversal through an electric field \cite{doi:10.1063/1.4984841}.
Following the same strategy, layered materials with formula $A_nB_nX_{3n+2}$ appeared to be other favorable candidates \cite{doi:10.1080/01411594.2014.986731, PhysRevLett.109.217202, Scarrozza2013}. 
In these crystal types, the octahedral rotations and the polar distortions are intrinsically coupled in a single mode through improper-like couplings too \cite{PhysRevB.84.075121}.
However, here again, while the proof of concept has been shown, a specific candidate is still missing.

In this letter, we show from first-principles calculations that within the family of the Barium-based layered fluorides BaMF$_4$~\cite{Eibschutz1969}, the case of BaCuF$_4$ has an ideal direct coupling between the polarization and the magnetization that does not rely on an improper mechanism.
Combined with the rather high N\'eel temperature T$_N$ = 275 K (the largest over $M$ = Mn, Ni, Co, and Fe series), this makes BaCuF$_4$ an appealing new candidate for electric-field-tuned magnetization. We also show that the peculiar coupling is linked to subtle interplay between ferroelectricity, magnetism and Jahn-Teller (JT) effect.

We performed density-functional theory (DFT) \cite{PhysRev.136.B864,PhysRev.140.A1133} calculations by using the \textsc{vasp} code (version 5.3.3) \cite{Kresse1996,Kresse1999}. The projected-augmented waves, PAW \cite{Blochl1994}, approach to represent the valence and core electrons was used. The electronic configurations taken into account in the pseudo-potentials as valence electrons are (5$p^6$6$s^1$, version 06Sep2000), (3$d^{10}$4$s^1$, version 06Sep2000), and (2$s^2$2$p^5$, version 08Apr2002) for Ba, Cu, and F atoms respectively. 
The exchange correlation was represented within the generalized gradient approximation GGA-PBEsol parametrization \cite{Perdew2008} and the $d$-electrons were corrected through the DFT$+U$ approximation within the Liechtenstein formalism \cite{Liechtenstein1995}.  
We used $U$ = 7.0 eV and $J$ = 0.9 eV parameter values that were optimized to reproduce the electronic band gap (3.9 eV) and magnetic moment of the HSE06 hybrid-functional calculation \cite{HSE, HSE06}.  
These parameter values are in good agreement with the ones used in KCuF$_3$ \cite{0953-8984-25-11-115404, Liechtenstein1995}. 
The periodic solution of the crystal was represented by using Bloch states with a Monkhorst-Pack \cite{PhysRevB.13.5188} \emph{k}-point mesh of 6$\times$4$\times$6 and 600 eV energy cut-off to give forces convergence of less than 0.001 eV$\cdot$\r{A}$^{-1}$. 
Spin-Orbit coupling (SOC) was included to consider non collinear magnetic configurations \cite{Hobbs2000}. 
Born effective charges and phonon calculations were performed within the density functional perturbation theory (DFPT) \cite{gonze1997} as implemented in \textsc{vasp}. The atomic structure figures were elaborated with the \textsc{vesta} code \cite{vesta}. 
Finally, the spontaneous polarization was computed by means of the Berry-phase approach \cite{Vanderbilt2000}.

In what follows, we start by describing the characterization of the ferroic orders in this compound to understand the correlation between the ferroelectricity and the weak-ferromagnetic state of BaCuF$_4$.

\begin{figure}[!t]
 \centering
 \includegraphics[width=8.5cm,keepaspectratio=true]{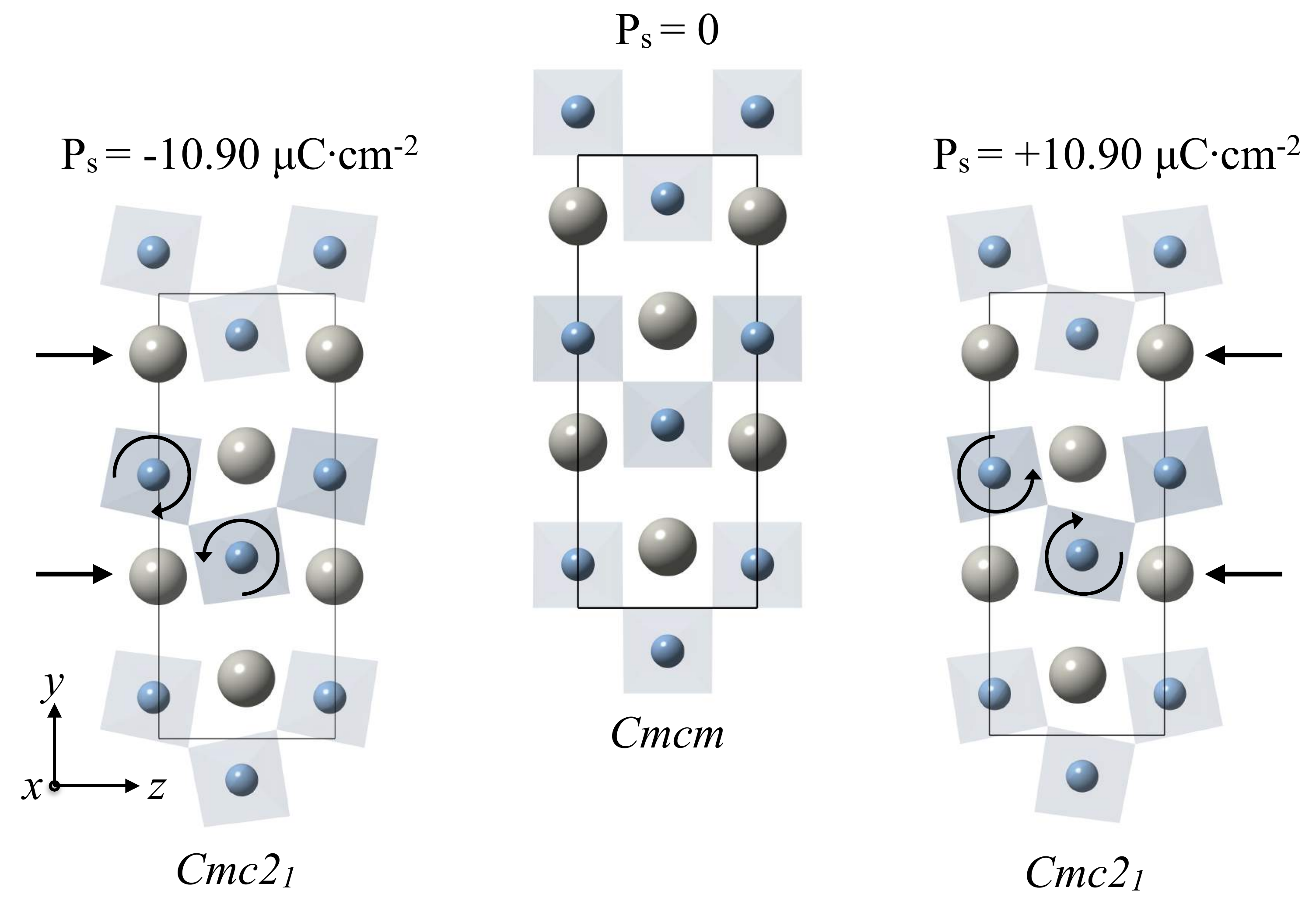}
 \caption{(Color online) $Cmcm$ and $Cmc2_1$ phases of the BaCuF$_4$. The phase transition responsible of the ferroelectric switching  involves the octahedral (in blue with Cu at the center) rotations around the $x$-axis as well as the polar Ba-sites (in dark grey). The later rotations and displacements, denoted with arrows, belong to the $\Gamma_2^-$ mode.}
 \label{F1}
\end{figure}


\emph{Ferroelectric ordering:}
The Ba$M$F$_4$ family of compounds is structurally characterized by octahedral $M$F$_6$ layers separated by Ba sheets stacked along the $y$-axis as shown in Fig. \ref{F1}.
We start by analyzing the phonon modes of the hypothetical high-symmetry structure of BaCuF$_4$ fluoride ($Cmcm$ space group, No. 63), not observed experimentally before the melting temperature (around 1000 K) \cite{DiDomenico1969}. 
The computed phonons dispersion of the high-symmetry reference (Fig.S1 in Supplementary Material \cite{sup-mat}) reveals the presence of three unstable phonon modes, when looking at the zone center and zone boundary points: $\Gamma_2^-$ at 66 $i$cm$^{-1}$, $S_2^+$ at 57 $i$cm$^{-1}$, and $Y_2^-$ at 47 $i$cm$^{-1}$. 

The $\Gamma_2^-$ mode is polar and combines polar motion of the Ba atoms along the $z$ direction with in-phase fluorine octahedra rotation around the $x$ direction (see Fig. \ref{F1}a) and its condensation reduces the symmetry of the crystal to the $Cmc2_1$ (No. 36) space group. 
The full relaxation of the crystal within the $Cmc2_1$ lowers the energy with respect to the $Cmcm$ phase by $\Delta$E = -27 meV per formula unit (meV/f.u.) and by decomposing the final distortions into symmetry adapted mode \cite{Orobengoa:ks5225, Perez-Mato:sh5107} we find a contribution of the $Cmcm$ $\Gamma_1^+$ and $\Gamma_2^-$ mode.
The $\Gamma_1^+$ mode is the mode that is invariant under all the symmetry operations of the $Cmcm$ phase, which means that it is a relaxation of the initial $Cmcm$ degrees of freedom that favors the development of the polarization.
A similar combination of modes has been reported in the ferroelectric LaTaO$_4$~\cite{LIU201631} and La$_2$Ti$_2$O$_7$ \cite{PhysRevB.84.075121}.

The $S_2^+$ mode drives the system to a non-polar phase with $P2_1/c$ space group (No. 14) and its eigenvector is an out-of-phase octahedra rotations around the $x$-axis (See Fig.S1 in Supplementary Material \cite{sup-mat}).
We note that it is the same type of distortion as for the $\Gamma_2^+$ mode (\textit{i.e.} octahedral rotation around the $x$-axis) but in the case of the $S_2^+$ mode, the out-of-phase octahedral rotations do not break the space inversion symmetry.
The gain of energy due to the relaxation of the $P2_1/c$ phase is $\Delta$E = -7 meV/f.u., thus about four times smaller than the $Cmc2_1$ phase.

The condensation of the $Y_2^-$ mode reduces the symmetry of the crystal into the $Pnma$ space group (No. 62) where its eigenvector involves in-phase clockwise rotation of the octahedra around the $x$-axis in one octahedral layer and an in-phase counter-clockwise in the next octahedral layer (Fig.S1 in Supplementary Material \cite{sup-mat}). 
The relaxation of the $Pnma$ phase lowers the energy by $\Delta$E = -10 meV/f.u., which is larger than the $P2_1/c$ phase but more than two times smaller than the $Cmc2_1$ phase.

We thus find that the ferroelectric $Cmc2_1$ phase is the ground state of BaCuF$_4$, which is in agreement with the experiments \cite{DANCE1981599, BABEL198577}. 
In this compound, in contrast with the other members of the same Ba$M$F$_4$ family, we note a strong JT distortion. The latter caused by Cu:$d^9$ orbital filling, which is also present in the high-symmetry $Cmcm$ structure and induces a large octahedra elongation along the $x$-axis. 
Therefore, we found that the Cu--F bonding distances in the $Cmc2_1$ phase are 2.25, 1.88, and 1.91 \r{A} for the bonds along the [1,0,0], [0,1,1], and [0,-1,-1] directions respectively.
The relaxed $a$, $b$, and $c$ lattice parameters are 4.453, 13.892, and 5.502 \r{A} respectively which are close to the experimental  values of 4.476, 13.972, and 5.551 \r{A} \cite{DANCE1981599, BABEL198577}. 
Thus, when comparing with other members of the family, the bonding Cu--F bonding length, along the x-axis, is by far the largest with an elongation close to 0.2 \r{A}. The later, induced by the strong JT-effect and having strong effects in the magnetic structure as will be discussed later.

\begin{table}[!b]
\caption{Polarization and ionic $M^{2+}$ radii \cite{Shannon1976} for all the members of the Ba$M$F$_4$ family. A clear trend of the polarization dependent of the octahedral cation size is observed in full agreement with the geometric proper ferroelectricity observed in these fluoride compounds. No experimental ferroelectric switching has been demonstrated for BaMnF$_4$ and BaFeF$_4$. Finally, in the last column are presented the energy barrier values for all the compound taken from the $Cmcm$ to the $Cmc2_1$.}
\begin{center}
\centering
\begin{tabular}{c c c c}
\hline
\hline
Compound &  $M^{2+}$ radii [pm]   & P$_s$ [$\mu$C$\cdot$cm$^{-2}$]   &  $\Delta$E [meV/f.u.]   \rule[-1ex]{0pt}{3.5ex} \\
\hline
BaNiF$_4$ &  83 &  6.8 \cite{PhysRevB.74.024102}   &  28 \cite{PhysRevB.74.024102}  \\
BaMgF$_4$ &  86 &  9.9   &    133  \cite{PhysRevB.93.064112} \\ 
BaCuF$_4$ &  87 &  10.9  &  27  \\ 
BaZnF$_4$ &  88 &  12.2   &  218   \cite{PhysRevB.93.064112} \\ 
BaCoF$_4$ &  88.5 &  9.0 \cite{PhysRevB.74.024102} &  58 \cite{PhysRevB.74.024102}\\
\hline
BaFeF$_4$ &  92 &  10.9 \cite{PhysRevB.74.024102} &  122  \cite{PhysRevB.74.024102}\\
BaMnF$_4$ &  97 &  13.6 \cite{PhysRevB.74.024102} &  191  \cite{PhysRevB.74.024102}\\
\hline
\hline
\end{tabular}
\end{center}
\label{tab:1}
\end{table}

The computed polarization in the ground state is $P_s = 10.9$ $\mu$C$\cdot$cm$^{-2}$, being in the range of amplitudes of Ba$M$F$_4$ compounds (see Table \ref{tab:1}). 
The computed energy difference between the $Cmcm$ to the $Cmc2_1$, is $\Delta$E = 27 meV/f.u., which is lower than the one reported for other family members such as BaMgF$_4$ and BaZnF$_4$ with barriers of 133 and 218 meV/f.u. but similar to BaNiF$_4$ and BaCoF$_4$ where the ferroelectric switching has been experimentally demonstrated \cite{Eibschutz1969}.
This low barrier value suggests that the ferroelectric switching can be easier. 
In order to estimate how the polarization of BaCuF$_4$ fits with respect to the members of the Ba$M$F$_4$ family, in Table \ref{tab:1} we compare the trend of the polarization's amplitude as a function of the $M^{2+}$ ionic radii \cite{Shannon1976}. 
We observe that the polarization follows the trend of the ionic radii size, which is expected from geometrically-driven polar displacements \cite{PhysRevB.89.104107}, also concluded from their close-to-nominal Born effective charges (see Supplementary Material \cite{sup-mat}) similarly as theoretically predicted \cite{PhysRevLett.116.117202} and later experimentally demonstrated in the multiferroic NaMnF$_3$ perovskite fluoride \cite{Yang2017}.
Interestingly, it can be also noted that BaMnF$_4$ and BaFeF$_4$, which lack of experimentally proved polarization reversal \cite{Eibschutz1969}, are those with the largest $M^{2+}$ ionic radii. 
The latter suggests a delicate balance between geometric effects and the switching process that needs to be further investigated.

\begin{figure}[!t]
 \centering
 \includegraphics[width=7.0cm,keepaspectratio=true]{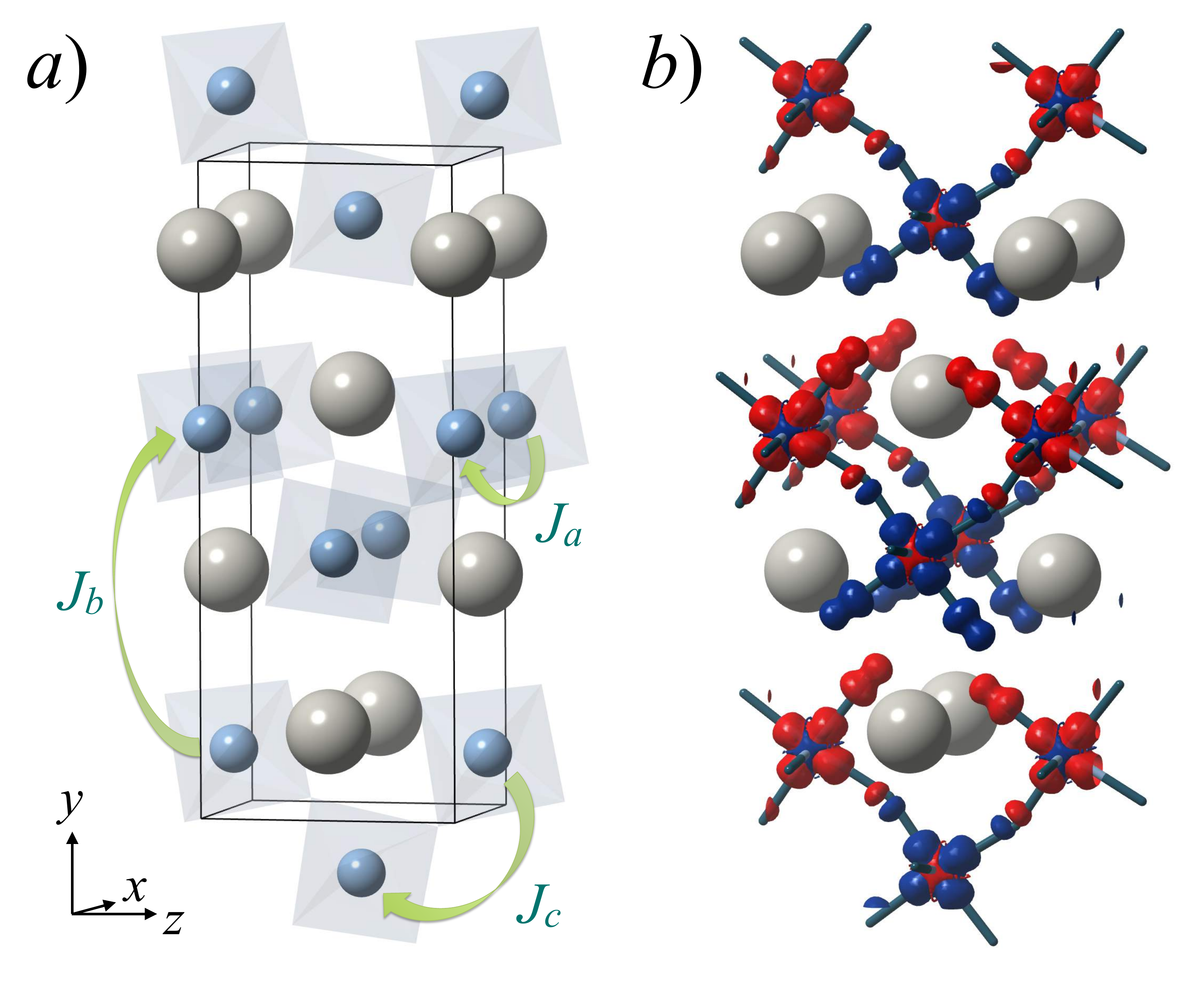}
 \caption{(Color online) $a$) $Cmc2_1$ structure where the magnetic exchange constants of the intralayer $J_a$ and $J_c$ and the interlayer $J_b$ constant are depicted. $b$ Spin-polarized charge-density where a clear orbital ordering, induced by the strong JT distortion, is observed. Here the magnetic moment up and down are depicted in red and blue colors respectively.}
 \label{F2}
\end{figure}


\emph{Magnetic ordering:}
Our analysis of the possible main collinear magnetic orderings of BaCuF$_4$ reveals the existence of an 1D-AFM thanks to its strong JT-distortion. 
This ordering is confirmed by the exchange constants (computed following the procedure of Ref. \onlinecite{PSSB:PSSB201451400}) where values of 0.04, 0.03, and -15.91 meV were obtained for $J_a$, $J_b$, and $J_c$ respectively (see notation in Fig. \ref{F2}$a$).
We find that $J_a$ and $J_b$ are very small which explains the 1D-AFM character at high temperatures. 
Moreover, the spin-polarized charge density (see Fig. \ref{F2}$b$), clearly shows the ferrodistortive character of the $d_{x^2-y^2}$ orbital ordering thanks to the 
strong JT-distortion present in this Cu: $d^9$ compound and then leading as a result to the 1D-AFM character.  
More details about the magnetic orderings can be found in the Supplementary Material \cite{sup-mat} and we would like to focus on the non-collinear magnetism analysis instead as next.

Starting from the 1D-AFM  (also known as $A$-AFM) in the non-collinear magnetic ordering regime, we observe the appearance of spin canting giving a weak-ferromagnetic (w-FM) canting along the $z$-axis, with $m_z$ = 0.059 $\mu_B$/atom. 
The system can be described by the modified Bertaud's notation \cite{bertaut, bousquet2016} as $A_yF_z$ where the $A$-AFM is the main ordering along the $y$-axis ($m_y$ = 0.829 $\mu_B$/atom) and canted ferromagnetism $F_z$ along the $z$-direction. 
We thus obtain a magnetic canting angle of about 4.12$^\circ$ with respect to the $y$-axis. 
Although, a non collinear ordering has been observed for $M$ = Mn, Fe, Co, and Ni too, the canted structure give rise to a weak-AFM ordering instead \cite{PhysRevB.74.024102, PhysRevB.74.020401} 
Then, our findings in the Cu case confirm that both, the spontaneous polarization and the ferromagnetic moment, are conveniently aligned along the $c$-axis.
Interestingly, this canting angle is larger than the those reported for the weak-AFM $M$ =  Ni \cite{PhysRevB.74.020401} and Ca$_3$Mn$_3$O$_7$ \cite{PhysRevLett.106.107204}. 
Besides, even when the magnetic measurements \cite{DANCE1981599} show that above 275 K the material exhibit a paramagnetic behavior, the JT-distortion is expected to remain in the structure due to a survival of the orbital ordering beyond T\textsubscript{N} as we observed in the relaxed $Cmcm$. 
This has a direct effect on the octahedral structure as observed for KCuF$_3$ \cite{PhysRevLett.101.266405}.

\begin{figure}[!t]
 \centering
 \includegraphics[width=8.0cm,keepaspectratio=true]{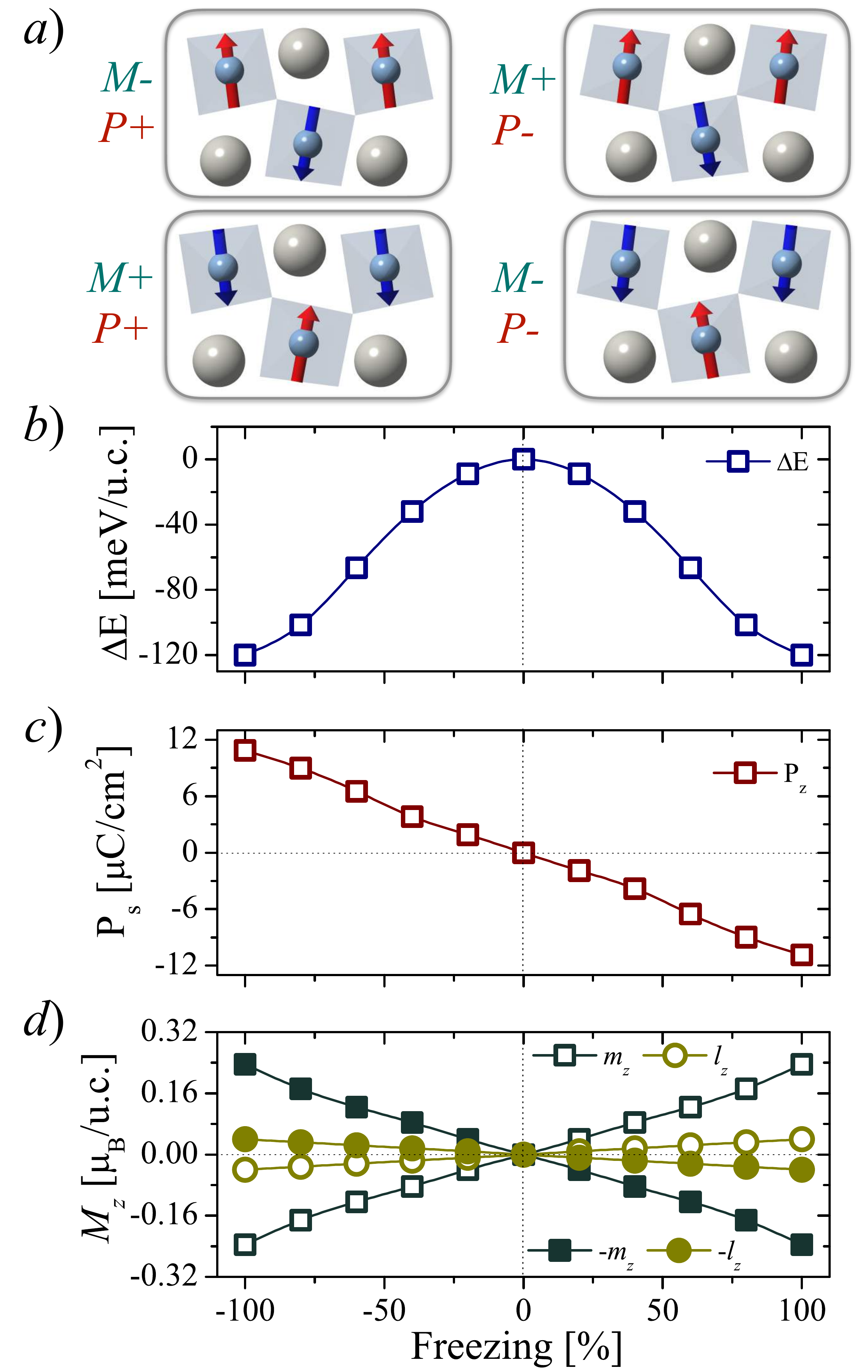}
 \caption{(Color online) $a$) Schematic representation of the four multiferroic states where the ferroelectric polarization and magnetization are shown in BaCuF$_4$. It can be observed that the magnetic moments, depicted as red and blue arrows for up- and down-orientation respectively, can be reversed by reversing the magnetic ordering and/or the polarization reversal. $b$) Double-well energy profile obtained thought the ferroelectric switching between the up and down spontaneous polarization orientations. $c$) Full polarization reversal going from $-$10.9 to $+$10.9 $\mu$C$\cdot$cm$^{-2}$. $d$) Magnetic ($m_z$) and orbital ($l_z$) moment, per u.c. reversal by means of the ferroelectric switching showing the correlation between the switching, rotations, and non collinear magnetic ordering.}
 \label{F3}
\end{figure}

\emph{Intertwined magnetization and polarization:}
Through the presence of both magnetization and polarization one can see that BaCuF$_4$ can hold up to 4 multiferroic states (\emph{i.e.} $M^-P^+$, $M^-P^-$, $M^+P^+$, and $M^+P^-$) as shown in Fig. \ref{F3}$a$). 
In the following, we are going to show that the the magnetization can be switched by means of an applied electric field, thanks to the spontaneous polarization reversal, and by an applied magnetic field thanks to its canted structure.

We performed the computed experiment where the full ground state distortion is gradually frozen into the $Cmcm$ reference phase. At each point the electronic structure is relaxed (and thus the non-collinear magnetization) with fixed structural parameters.
In Fig. \ref{F3}$b$ we show the energy well and the associated polarization at different amplitude of distortions in Fig. \ref{F3}$c$.
The ferroelectric polarization shows a full reversal from $-$10.9 to $+$10.9 $\mu$C$\cdot$cm$^{-2}$.
The spin ($m_z$) and orbital ($l_z$) moments, plotted in Fig. \ref{F3}$d$,  shows that $m_z\gg l_z$, and most importantly that the magnetization, changes its sign when the polarization is reversed. 
Therefore as expected, the four multiferroic states are possible, and more importantly, their magnetization and polarization directions  could be reversed. The later showing that a proof of concept of a 4-states memory is possible in this type of materials.

This link between the polarization and the magnetization can be explained in terms of the Dzyaloshinskii-Moriya (DM) interaction \cite{Dzyaloshinsky, Moriya} that is related to the rotation of the octahedra. 
The DM interaction energy is defined by the relationship E = \textbf{D$_{ij}$}$\cdot$($s_i$ $\times$ $s_j$), where \textbf{D$_{ij}$} is the DM tensor and the $s_i$ and $s_j$ are the spins related to the ions $i$ and $j$ respectively. 
As demonstrated in perovskites \cite{Hyun2011}, the DM tensor can be related to the inter ionic vectors as \textbf{D$_{ij}$} $\propto$ ($\hat{x}_i$ $\times$ $\hat{x}_j$) \cite{Hyun2011}, in which $\hat{x}_i$ and $\hat{x}_j$ are unitary vectors along the Cu--F--Cu bonds. 
In BaCuF$_4$ the absence of octahedral rotation in the high-symmetry $Cmcm$ structure gives a 180$^\circ$ Cu--F--Cu bonding angle and thus forbids a weak-FM ordering. 
A key feature in this type of systems is that the polarization and octahedra rotations are embedded into the same unstable mode of the $Cmcm$ structure such that, reversing the polarization will systematically reverse the octahedral rotation and thus the weak-FM.
It is important to mention that the non collinear ordering is also observed in all of the other magnetic phases ($G$-AFM and $C$-AFM) but never with a weak-FM moment.

BaCuF$_4$ ideally combines the desired magnetic ordering, thanks to the JT-distortion, with the appropriated structural ground state over the $AM$F$_4$ family to exhibit a perfect electric-field magnetization reversal. It has thus a large magnetoelectric effect at rather large temperature and can be used to build a 4-states memory device as discussed before.

Although at first glance the magnetic moment could be seen to be weak, it could be amplified by layer-engineering and growing a ferromagnetic layer on the top of it as demonstrated through the exchange-bias effect  \cite{Matsukura2015} in [Co/Pd(Pt)]/Cr$_2$O$_3$ \cite{PhysRevLett.94.117203} and NiFe/h-YMnO$_3$ (LuMnO$_3$) \cite{PhysRevLett.97.227201} but also in the Barium-based family of $AM$F$_4$ crystals with $M$ = Ni and Mn \cite{Zhou2015, Zhou2017}. 
Therefore, bilayered EB-effect could be combined to bring about a novel electrically-controlled magnetic system based-on BaCuF$_4$  \cite{Matsukura2015}.
In conclusion, we believe the BaCuF$_4$ compound is an ideal candidate to show a strong multiferroic/magnetoelectric coupling close to room-temperature, being to our knowledge, the only fluoride material that exhibits such behavior close to room-temperature \cite{Scott2011a}.


\begin{acknowledgments}
\emph{Acknowledgements:}
This work used the XSEDE which is supported by National Science Foundation grant number ACI-1053575. ACGC and EB acknowledge the ARC project AIMED and the F.R.S-FNRS PDR project MaRePeThe. The authors also acknowledge the support from the Texas Advances Computer Center (with the Stampede2 and Bridges supercomputers),  the PRACE project TheDeNoMo and on the CECI facilities funded by F.R.S-FNRS (Grant No. 2.5020.1) and Tier-1 supercomputer of the F\'ed\'eration Wallonie-Bruxelles funded by the Walloon Region (Grant No. 1117545).
This work was supported by the DMREF-NSF 1434897, NSF OAC-1740111 and DOE DE-SC0016176 projects.
\end{acknowledgments}

\bibliography{library}

\end{document}